# Induction of tin pest for cleaning tin-drop contaminated optics

Norbert Böwering[a,b*]

[a]*BökoTech, Ringstraße 21, 33619 Bielefeld, Germany* and

[b]*Molecular and Surface Physics, Bielefeld University, 33501 Bielefeld, Germany*

**Highlights**

- Tin pest induction leads to embrittlement of tin drops after β → α Sn transformation.
- The purity grade influences the transformation speed of tin very strongly.
- Tin drops on multilayer-coated optics disintegrate after β → α Sn transformation.
- Tin drop contamination of optics is cleaned via phase transformation at -24 °C.
- Reflectance of multilayer-coated mirrors is restored after tin drop transformation.

**Graphical Abstract**

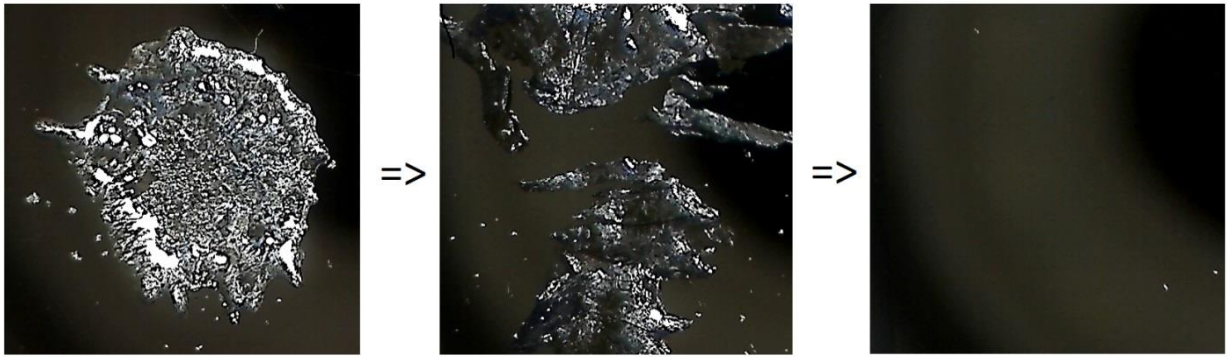

**Abstract**

Tin pest, the allotropic β → α phase transformation of tin, was examined for use in cleaning of tin-contaminated optics. Induction of change in material structure led to disintegration of tin samples into pieces and powder. The transition times were studied for tin drops of different purity grades, using inoculation with α-Sn seed particles, also after prior mechanical deformation and surface oxide removal. For tin of very high purity levels fast nucleation within hours and full transformation within a day could be achieved during cooling at -24 °C, resulting in strong embrittlement of the material. Tin dripped onto samples of multilayer-coated optics as used in extreme ultraviolet lithography machines was made cleanable by phase transition after inoculation and cooling. The reflectance of multilayer-coated mirrors was found to decrease by no more than 1% with this cleaning method.

\* Corresponding author

*E-mail addresses:* boewering@physik.uni-bielefeld.de  boekotech@web.de (N. Böwering).

§*Abbreviations:* LPP: laser-produced plasma, EUV: extreme ultra-violet, MLM: multilayer mirror, CCD: charge-coupled device.

1. **Introduction**

Tin pest is the name of the structural phase transformation of the ductile metal *white tin* (β-Sn) to the brittle semimetal *gray tin* (α-Sn) [1-5]. It can take place at temperatures below the transition temperature of 13.2 °C [1], where α-Sn becomes the stable phase of tin with lower entropy compared to β-Sn, which is then in principle metastable. However, even when cooling a bulk sample of pure white tin to negative Celsius temperatures for long periods, this allotropic phase transformation occurs rarely or is very difficult to induce unless certain prerequisites are met. These include inoculation seeding with α-Sn, removal of the tin oxide surface layer or prior deformation of the sample [2-8], as discussed in detail further below. On the other hand, once small centers of gray tin have indeed started to nucleate on the surface of the material, the subsequent phase transformation takes place fairly readily if the specimen is held at cold temperatures in a range of about -20 °C to -50 °C [1-3, 8]. In a displacing rearrangement of the tin atoms in the solid matrix, white β-Sn with body-centered tetragonal crystal structure transforms into gray α-Sn with face-centered cubic structure. Since this structural change of the material is also accompanied by a large increase in volume of about 26%, the resulting product, gray tin, shows an extreme brittleness and a strong tendency to crack or blister so that it can easily disintegrate into pieces or powder [1-3, 7].

In past basic research structural and semiconducting properties of the semi-metallic phase α-Sn were examined in detail [1, 9-11]. Alloys of tin and lead that were used for many years in soldering applications were found to be not very prone to tin pest [5, 12]. However, with the recent global introduction of lead-free solders in the electronics industry to replace formerly used Sn-Pb alloys, the potential susceptibility to tin pest at low temperatures has attracted renewed vivid attention. The topic was re-investigated by several groups for typical tin-rich alloys used in electronic soldering applications with the goal of better understanding the factors influencing the start and spread of tin pest and its dependence on alloy components and impurities [5, 8, 11-18].

Generally, in the past the focus of many studies of the β → α tin allotropy was on better comprehension in view of the prevention of tin pest occurrence in order to avoid undesired outcomes or catastrophic consequences originating from the spread of gray tin leading to unwanted rapid material disintegration. It was found that minor impurities in the tin specimens can influence the transformation strongly. Minute concentrations of impurities like Sb, Bi and Pb, each element having an appreciable solid solubility in the tin matrix, can suppress or slow down the transition to gray tin significantly [1, 8, 12-17], since they hinder lattice expansion by the pinning of dislocations.

In contrast to these studies, in this communication a new emerging application in the context of optics cleaning is described where the fast initiation of tin pest is indeed the desired outcome. The purpose of this work is to examine the transformation of tin-contamination on drops of previously untransformed β-Sn that can attach to the surface of highly sensitive optics used for example in machines for semiconductor manufacturing. Here, the aim is to investigate if and under what conditions the allotropic phase transition to α-Sn can be induced with advantage in order to transform solid tin drops into brittle pieces that can easily be removed from the optic without any damage. The transformation can then lead to a sizable reduction of tin deposition. This is often rather difficult to achieve in reasonably short time by other methods of dry cleaning like, for example, plasma etching schemes with hydrogen radicals [19, 20].

High-purity tin droplets are used as the target in laser-produced plasmas (LPP)$^§$ that have been developed as light sources at extreme ultraviolet (EUV)$^§$ wavelengths near 13.5 nm for next-generation semi-conductor production by lithography technologies [21-24]. The emitted EUV light from the hot tin plasma is collected and shaped by sensitive reflective optics. In the EUV source the LPP is generated by focusing a laser beam on tin droplets that pass in a stream directly in front of the large light-collecting mirror [21-24]. This collector with incidence angles near normal has a multilayer mirror (MLM)$^§$ coating of many alternating Mo and Si layers with a typical bilayer spacing of about 7nm to provide Bragg reflection of the incident EUV light from the tin plasma towards the illumination optics of the EUV scanner [25, 26].

LPP generation is accompanied by tin debris impinging on the collector mirror and other surrounding surfaces leading to accumulating tin contamination [24, 25]. Debris mitigation schemes based on flows of hydrogen gas in the source are only partially effective at the low pressure environment required to transmit EUV radiation [22]. Hydrogen radicals generated by absorption of EUV radiation can transform thin tin deposits on the mirror to volatile $SnH_4$ (tin tetra-hydride) molecules that are pumped away, resulting in a net self-cleaning effect as long as the tin deposition rate on the optical surface is not too high [24]. Furthermore, in-situ cleaning methods using a source of hydrogen radicals have been described for the removal of tin layers with thickness of up to tens of micrometers [22, 27]. Thicker and more massive material depositions cannot be removed quickly by in-situ etching alone. Application of expensive special cleaning and ex-situ refurbishment techniques to contaminated optics are then required [28]. The optics most prone to contamination are the EUV collector mirror and the last deflection mirror of the incident laser beam which are located closely below the plasma [21, 24]. Tin deposits on the chamber walls can be melted by absorption of laser radiation and drop down onto the collector mirror sticking to its surface thus reducing its EUV-reflective area.

For cleaning, mechanical removal is not an option when a tin drop adheres well to a mirror surface since such action can remove and damage (part of) the coating. Chemical wet etching of tin or top MLM layer stripping may be a possibility for cleaning by refurbishment [28, 29], but it is generally associated with lengthy disassembly and re-assembly of the entire collector module. Avoiding disassembly of the contaminated EUV optics, dry cleaning methods using gaseous or frozen-particle flows, e. g. $CO_2$ (carbon dioxide) snow pellets as described for the cleaning of an EUV light source or mask, may be applied [30, 31]. Such techniques are most suitable for the removal of rough and powdery layers and not effective for thick smooth deposits. Therefore, it would be advantageous to induce in fairly short time a phase change of β-Sn drops or other thick β-Sn deposits into brittle α-Sn powder in order to increase the effectiveness of other removal techniques that may be applied subsequently. With this concept in mind, here the conditions are examined for transformation of white β-Sn to gray α-Sn pieces and powder both for loose drops created by the dripping of tin granules onto a metal plate and also for tin drops on samples of MLM-coated EUV optics.

## 2. Experimental

The samples used in this work were previously untransformed drops of pure beta tin. Flat drops of typically around 10 mm diameter were either obtained directly from a supplier (at 99.9% purity) or generated (in cases of higher purity grades) from β-Sn granules with a mass of 0.2 - 0.4 g by dripping in air onto a metal plate. For tin dripping, a heated copper plate (1 mm thick, temperature ~250 °C) was used with a 3 mm diameter hole on which the tin specimen was placed and through which it fell after melting, subsequently hitting a 3 mm thick stainless steel plate with smooth surface located below at a distance of 30 cm. Small vibrations were induced at the Cu plate to facilitate the release of the tin drop from the plate soon after melting. Generally, the oxidized tin drops produced in this way, solidified after impact, had a flat round disc shape showing variations in local thickness in the range of 0.1 – 1.0 mm, typically with largest thickness occurring at the rim, often with a "finger" pattern at the edge. The calculated droplet speed at impact is 2.4 m/s. The generated tin drops did not stick to the stainless steel surface and could be used as samples for testing of the conditions that induce the allotropic transformation.

The nominal purity grades of the bulk tin used for the examined samples (labeled A – E) are listed in Table 1. For the lowest purity grade of tin used here, Sn type A, the *maximum levels* of relevant impurities, as listed by the supplier (Westfalenzinn), are: 400 ppm Pb, 300 ppm Sb, 100 ppm Bi, 300 ppm As, 300 ppm Fe, 100 ppm Cu [32]. However, according to supplier information trace element levels in *typical* samples amount to generally a fraction of less than ~1/3 of these *maximum* numbers, except for Pb, leading to a *typical* total purity level of about 99.96%. For purity grades of 99.999% the impurity numbers are expected to be correspondingly 100 times smaller, in the range of only a few ppm or even less. Analysis provided by the suppliers for typical samples corresponding to Sn type D and E of Table 1 indicate that the expected combined contributions of trace elements Pb, Sb and Bi amount to <5 ppm and <1 ppm, respectively.

**Table 1:** Nominal purity grades of used tin samples from different suppliers

| Type | Purity grade (%) | Supplier | Description |
|---|---|---|---|
| A | 99.9 | Westfalenzinn | melted from pure tin anodes |
| B | 99.9 | GTS-Siliton | guaranteed minimum purity |
| C | 99.99 | Metoxleg | high purity |
| D | 99.999 | Fine Metal Corp. | industrial grade high purity |
| E | 99.999 | Honeywell | research grade high purity |

Cooling of all samples, each stored in small transparent plastic sample containers, was carried out at ambient atmosphere in the freezer of a refrigerator at a constant temperature of -24 °C. This temperature is sufficiently low for nucleation and fast growth rate of already nucleated α-Sn. It is close to conditions of maximum growth velocity which was found by several groups to occur at around -35 °C, nearly independent of the grade of high-purity tin [1-3, 11, 12]. For detection of α-Sn visual observation of color changes from shiny metallic luster of β-Sn to gray diffuse light reflection of semiconducting α-Sn was used. Inspection of samples by brief removal from the freezer was done at room temperature in regular intervals. Inspection and recording was made by using a camera with charge-coupled device (CCD)[§] sensor or with a reflected light microscope with attached light-emitting diode illumination and CCD camera. When required, contactless temperature measurements were made with an infrared thermometer. The experimental process is shown schematically in the diagram of Fig. 1.

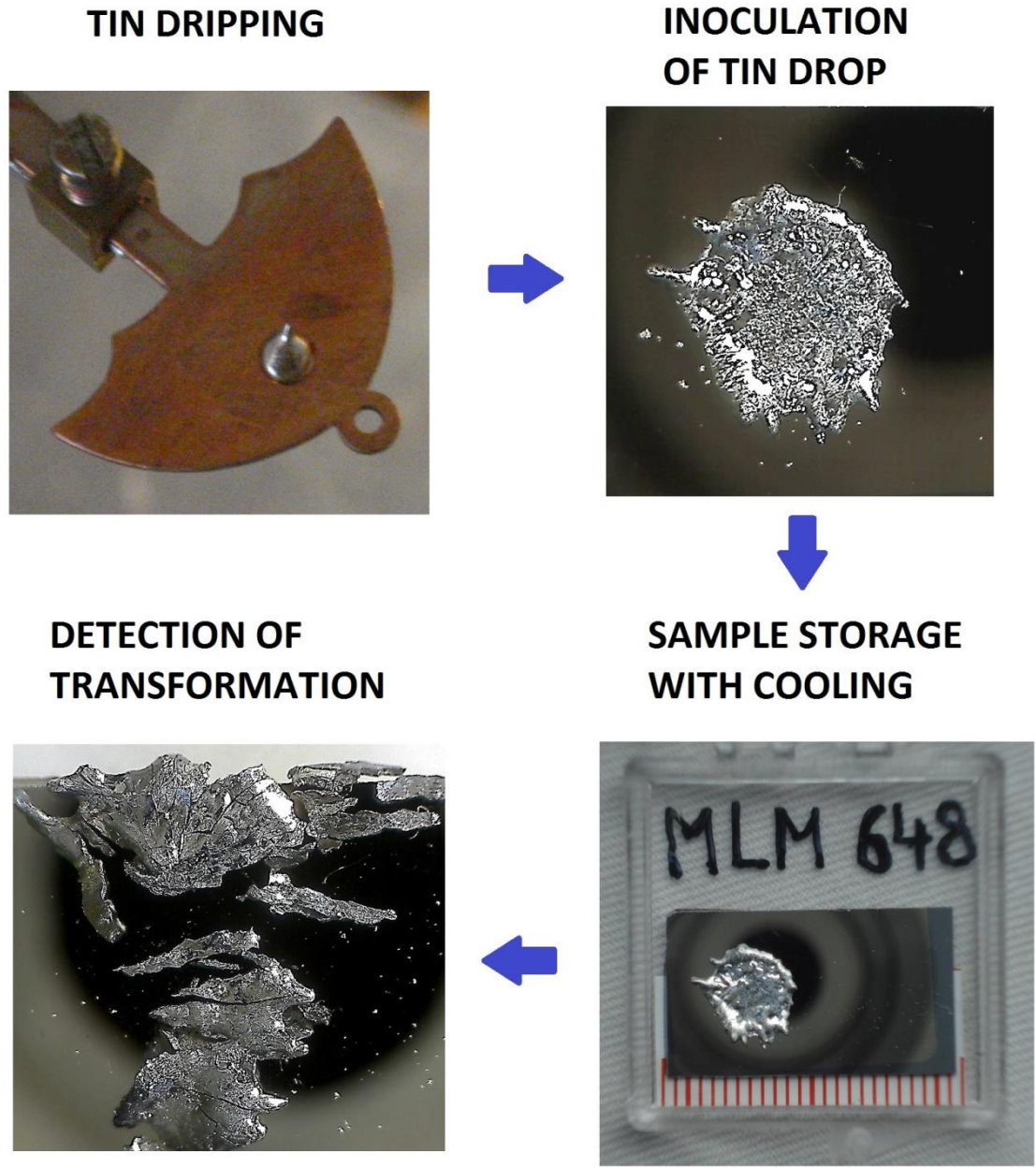

**Fig. 1.** Schematic process diagram representing the main experimental steps: tin dripping, inoculation treatment, sample storage with cooling, detection of transformation.

In a general description the phase transformation of β-Sn to α-Sn involves initial nucleation on the sample surface and subsequent growth of gray tin regions on the surface and into the bulk [1-3, 5, 12, 15-18]. The process is often very sluggish, at least for previously untransformed white tin, unless special means are employed to avoid long nucleation times. In this work several techniques were applied in order to increase the speed of the allotropic transformation strongly. Inoculation seeding (see also [1-4, 7, 8, 17]) with small-size α-Sn seed powder was generally used for initiation of the autocatalytic conversion at several locations on the sample surface. Seed powders of small grain size α-Sn were produced in a porcelain jar by several successive transformations between the beta and alpha phase of tin by repeatedly cycling with alternating prolonged cooling (to -24 °C) and short heating (to above 60 °C). Seeds were very gently pressed in or just put in surface contact with the tin drop using clean plastic tweezers. Inoculation with small pieces of isomorphic crystals from a high-purity CdTe wafer and with fine Ge powder (99.999% purity) as foreign seeds [1, 3, 4, 8] was initially also tried but did not yield any faster nucleation rates compared to what was obtained with α-Sn seeds; this method was therefore no longer pursued. The ~1nm-thick thin oxide layer ($SnO_2$) that develops on tin surfaces during prolonged exposure to air [33] can prevent direct epitaxial contact with the seed crystals. Consequently, drops were often (but not always) dipped in <24% concentrated HCl (hydrochloric) acid solution for a few minutes, followed by brief rinsing in de-mineralized water, in order to remove surface oxides before subsequent inoculation with seeds (see also [1-8]). The polycrystalline material structure of the samples became apparent after removal of the surface oxide layer by the acid.

It is known that prior mechanical deformation can increase the speed of α-Sn nucleation substantially [1-8, 14, 17]. In particular, a pronounced increase was observed in studies where less pure tin (~99.9%) was used [8, 13, 14, 16, 17]. The mechanism is attributed to lattice distortions after deformation leading to increased dislocation density and stored internal strain energy which can facilitate the allotropic β→α transition [3, 5, 8, 14, 17]. The transformation speed can be increased significantly by prior cold-working even at room temperature although recrystallization and recovery processes are then already expected to occur [3]. Specimen of white tin with lower purity (types A, B and C) were deformed at temperatures of 17 °C - 19 °C to reach a typical thickness reduction of 50% or more. For mechanical deformation the samples were wrapped in Al foil and rolled out using a stainless steel rod for pressing.

For testing of EUV optics small rectangular and square pieces of Si (100) wafers (up to 1" size) with the surface polished to <0.2 nm rms roughness were used. They were multilayer-coated by electron-beam evaporation, deposited at Bielefeld University [34], and magnetron sputtering, deposited by optiX fab [35], respectively. The MLMs consisted of uncapped Mo/Si multilayers (50 bilayers) optimized for a peak reflectance $R_{max}$ above 0.65 at 13.5 nm at near normal incidence. For exposure to tin dripping they were placed at 30 cm distance on the metal plate below the heated Cu drip plate. The EUV reflectance of a sample from optiX fab was measured at PTB Berlin in a reflectometer using synchrotron radiation [36].

3. Results and discussions

A consistent uniform interpretation of tin pest transformation based on results of past studies for pure tin [2-4, 6, 7] and tin alloys [5, 8, 12-18] is difficult to obtain due to large variations of the different individual test conditions. Therefore, in order to reach a better understanding of the requirements for fast induction of tin pest, isolated tin drops were studied first without a substrate. Initially, tests were carried out with β-Sn type A (99.9% purity grade) using surface inoculation with α-Sn seed particles. However, no phase transformation could be induced in reasonable time (on the order of several weeks) on untreated tin drops cooled at temperatures of -24 °C even when the seeds were pressed into the specimen. On the other hand, when in addition mechanical deformation with ~50% thickness reduction and surface oxide removal with HCl was applied prior to inoculation with α-Sn seeds, the onset of tin pest growth in small regions could be observed just within several days after sample storage at -24 °C by the change from shiny metallic to dull gray color. Full completion of the transformation across the surface of the drop required a considerably longer time. A similar behavior was found during use of tin drops of Sn type B with same purity grade, but obtained from a different supplier (see Table 1). These results with slow transformation induced by the combined use of seeding, mechanical deformation and surface oxide removal are comparable to the recent observations reported by Skwarek et al. for the case of alloys of 99% Sn with 1% Cu using tin of similar purity level [8].

For tin drops of higher purity grade 99.99% (Sn type C) the start of tin pest could be induced after several days of cooling even when no prior deformation was made. Fig. 2 shows the general development of the phase transformation observed on a representative typical tin drop sample using Sn type C. The surface of the tin drop was treated with HCl and then inoculated with a number of α-Sn seed particles mainly to the left of the central region (Fig. 2a). After 9 days storage at -24 °C gray tin blisters had spread in the infected region covering about 5% of the total surface area, as shown in Fig. 2b. With continued cooling the regions of gray tin grew further and merged. After 32 days of cooling nearly half of the surface area was transformed (Fig. 2c). Due to large volume expansion the sample was also strongly deformed and a crack started to open up at the center. The tin drop then began to disintegrate and broke up more and more as the transformed surface area increased further (see Fig. 2d). When the sample was fully converted it became very brittle and was easily broken into pieces, even when touched only gently with plastic tweezers. Evidence of phase transformation is from the color change from shiny silvery color to dull gray, from the observed deformation of the drop due to strong volume expansion, and from the transition from ductile to highly brittle material behavior revealing the structural change.

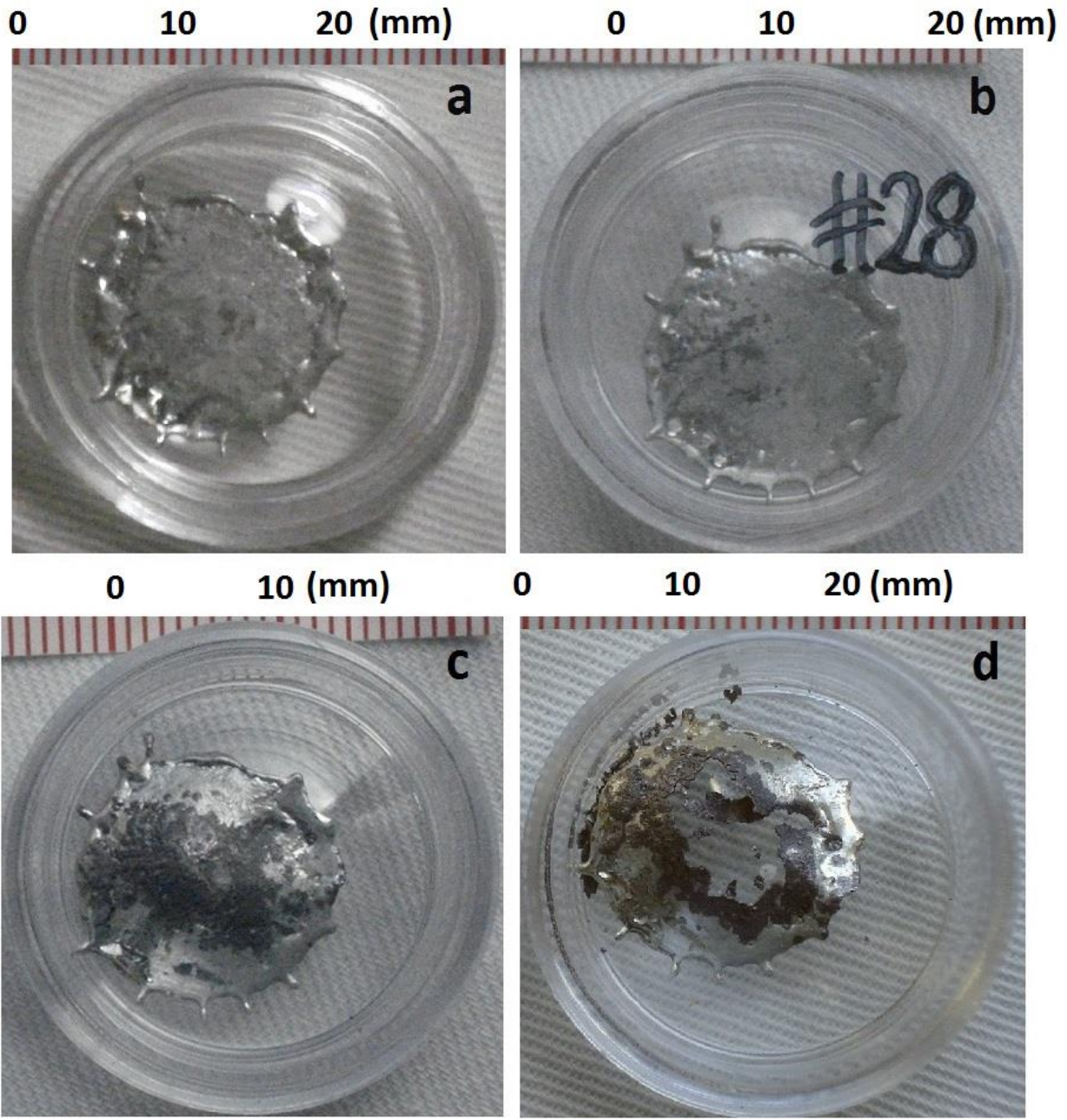

**Fig. 2.** Evolution of tin pest on a tin drop stored at -24 °C using tin with purity grade 99.99%: a) after inoculation, b) appearance after 9 days of cooling, c) after 32 days of cooling, d) after 42 days of cooling.

Tin drops of Sn type D and E (purity grade 99.999%) showed in principle a similar behavior, but with two significant differences: In contrast to the lower purity grades the observed transformation to gray tin was much faster (taking now hours instead of days); when large regions of α-Sn were formed they had a stronger tendency to develop long cracks. Generally, samples of Sn type E transformed even faster than those of type D, showing first signs of infection as soon as one hour after start of cooling. The high transition rates can be attributed to the very high purity grades of the material, with only very low amounts of trace elements Pb, Sb and Bi being present. Their inhibiting influence on the allotropic transformation was noted in previous studies [1, 12-16]. As described in Sec. 2, Sn type E is expected to have even lower levels of impurities compared to type D.

Inoculation by contacting the surface with seed particles was found to be a prerequisite for initiation of nucleation of α-Sn on all tin drops examined here. As discussed by Styrkas [4], in principle other mechanisms can also stimulate the nucleation of α-Sn on previously not converted β-Sn samples, but generally they require considerably longer initiation times. It was found here that prior deformation and even surface oxide removal was not an essential ingredient for fast transformation (on the order of 24 hours) of drops with a tin purity grade of 99.999%. Even when seeds of α-Sn were just put in contact with the surface of the tin drop, a full conversion to gray tin could typically be achieved in less than 12 hours for Sn type E, and within at most 22 hours for Sn type D. When the surface oxides were removed in addition by dipping in HCl acid, samples of β-Sn of type E showed generally a full conversion to gray tin in less than 8 hours. Similarly fast transformation times were recently reported by Ref. [7] for specimen with comparably high degree of tin purity.

The behavior of inoculated tin drops was compared for different treatments and purity grades (listed in Table 1) in order to obtain a clear overview and baseline on the time development of the phase transition under different conditions. The relative surface area coverage of converted α-Sn as a function of cooling time at a temperature of -24 °C was measured by evaluation of photographs taken during inspection for phase transformations of drops with different purity grade, generally examining several samples of each type. The treatment before inoculation with respect to deformation and oxide removal was also varied. In Fig. 3 the evolution of the relative area coverages for some representative samples are compared on a logarithmic time scale. As described, samples of type E (round symbols) showed the fastest conversion when their surface was also treated with HCl acid. Samples of type D (triangles) exhibited a slightly slower kinetic behavior compared to type E, but showed also slightly faster conversion when the surface oxides were removed before infection. For the samples of tin type E on average a gray tin growth rate of (1.1 ± 0.1) mm/h was determined. This is in agreement with previous studies where linear growth rates of α-Sn close to 1 mm/h were found for tin with purity levels of 99.999% or higher [2, 3]. However, the conversion proceeded significantly more slowly for lower purity grades of tin. Tin drops of type C (square symbols in Fig. 3) dipped in HCl before inoculation resulted in considerably longer conversion rates, leading to a full transformation of the tin drop to α-Sn only after more than 2000 hours. Times of roughly 1000 hours were needed for transformation when the samples of type C were in addition also mechanically deformed. To increase the transformation rate, prior deformation with ~50% thickness reduction was also made for the case of the nominally less pure samples of Sn type A and B (diamonds and crosses in Fig. 3). Still, close to ~1000 hours of cooling at -24 °C, or even more were required for achieving a full conversion to α-Sn. Samples of type B purity grade were generally found to have the slowest conversion rate. Two data sets each for tin of type E and C shown in Fig. 3 were obtained under the same conditions. They are shown to illustrate the typical variations observed in the measurements which can be attributed mainly to differences in initial incubation time.

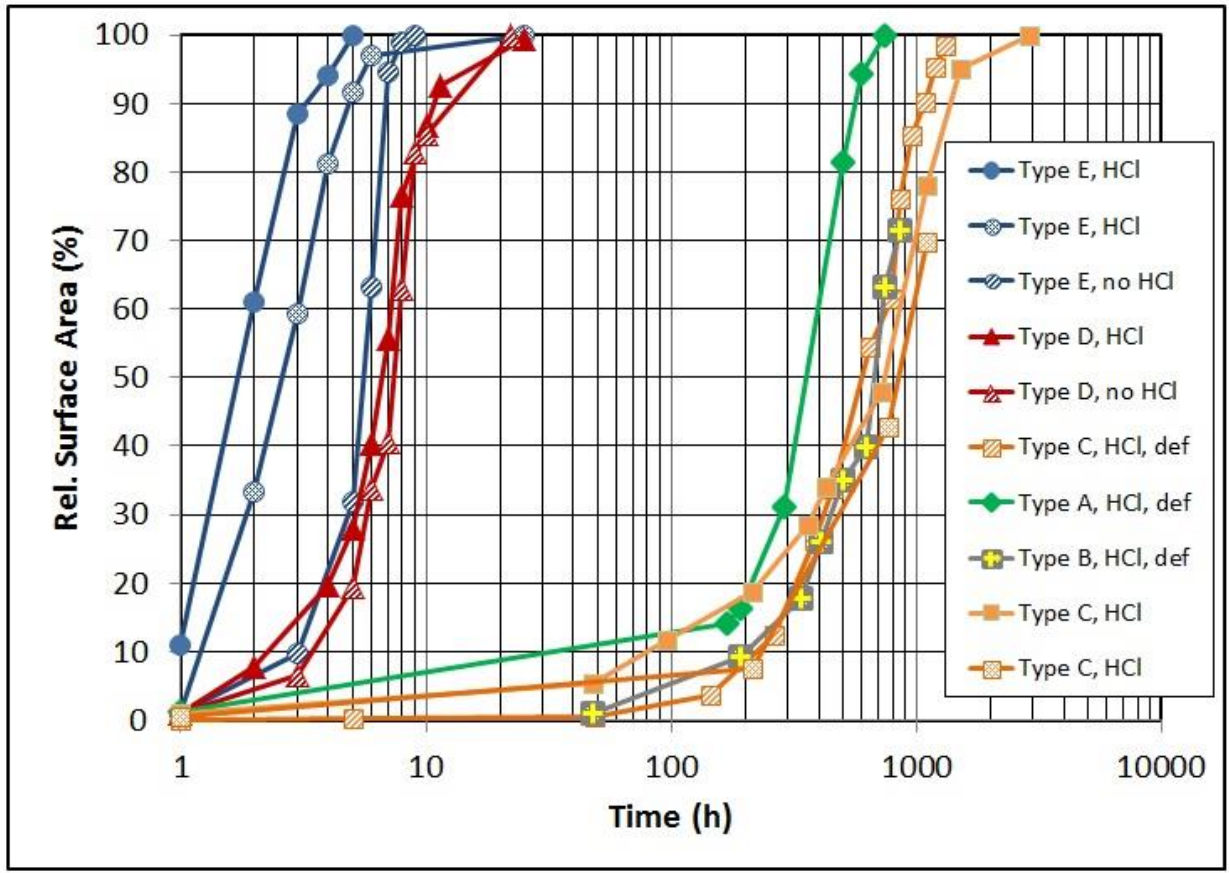

**Fig. 3.** Time evolution of surface area converted to α-Sn during storage at -24 °C for tin drops of purity grades A - E and for different treatment. Full and shaded circles: Sn type E, dipped in HCl; dashed circles: Sn type E, not treated with HCl; full triangles: Sn type D, dipped in HCl; dashed triangles: Sn type D, not treated with HCl; full diamonds: Sn type A, dipped in HCl and deformed; full crosses: Sn type B, dipped in HCl and deformed; dashed squares: Sn type C, dipped in HCl and deformed; full and shaded squares: Sn type C dipped in HCl, not deformed.

For testing of the cleaning method tin drops were produced on top of multilayer-coated optic samples by the tin dripping scheme described in Sec. 2. A drop of Sn type C was dripped on a piece of MLM-coated Si wafer. After local removal of surface oxides by use of HCl, the drop was treated with α-Sn seed particles and started to transform slowly into α-Sn during cooling. However, in this case it took close to three months of storage at -24 °C for reaching nearly full allotropic conversion of the tin material. Therefore, for further tests only β-Sn of the highest available purity levels D and E was used. This resulted in much faster conversion times. Fig. 4 shows a magnified view of the observed conversion of a tin drop after dripping on a MLM-coated Si-wafer sample during cooling using tin of type E. The inoculated drop (Fig. 4a) developed a blister of gray tin which buckled and deformed in shape due to volume expansion, covering an area of about 1/3 of the drop after about 4.5 hours (Fig 4b). It broke open in the center and led to full structural transformation with disintegration into gray pieces and some powder in less than 22 hours, when the next photo was made (Fig 4c).

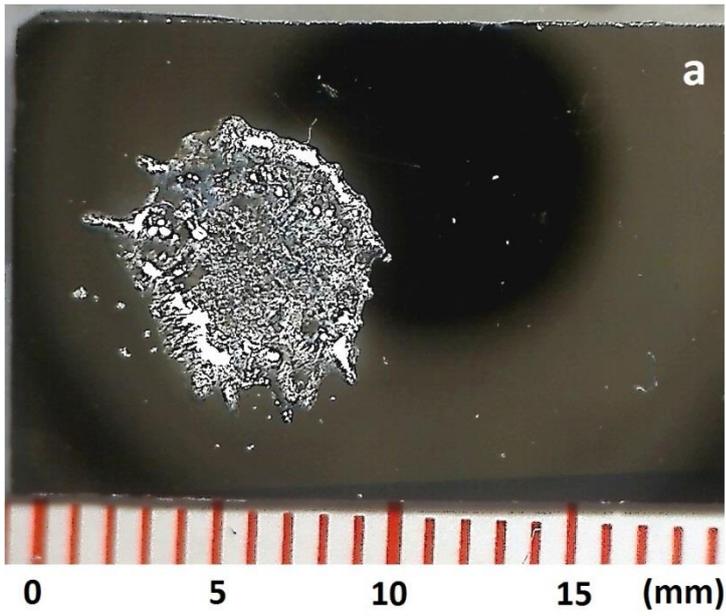
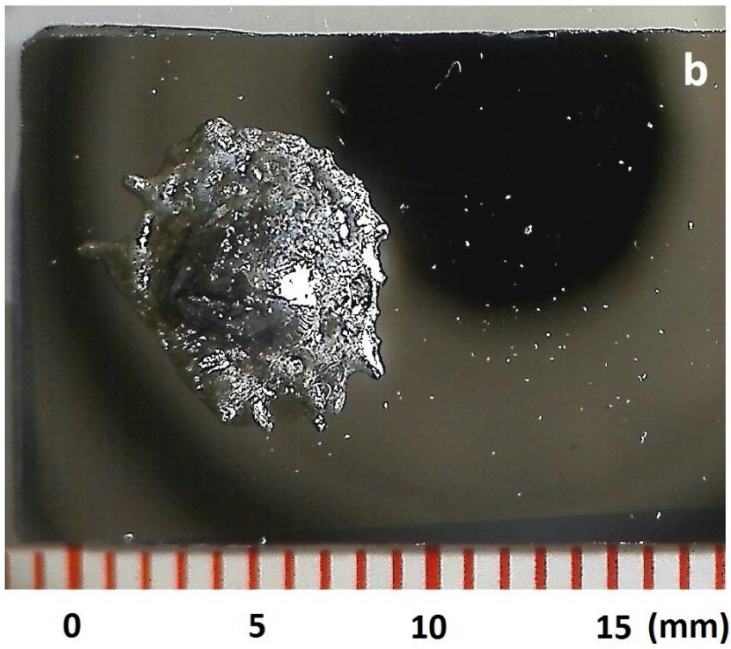
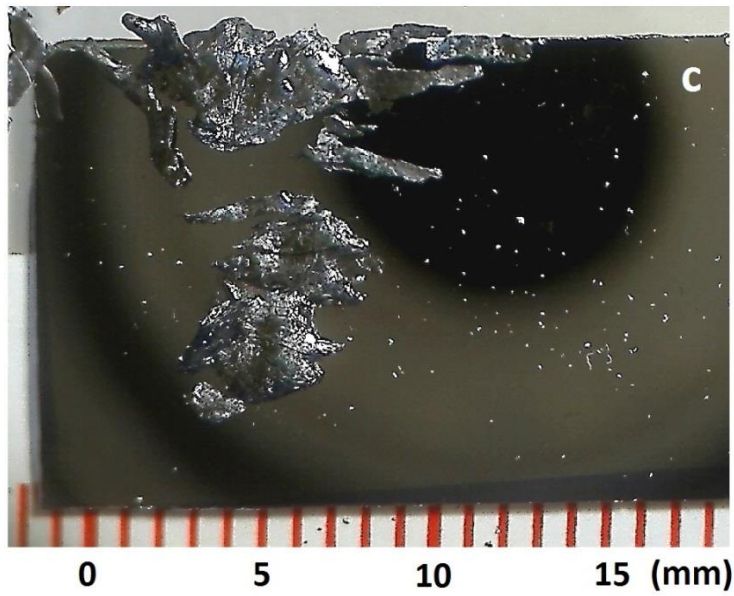

**Fig. 4.** Phase transformation of tin drop of 99.999% purity on Mo/Si-coated MLM during storage at -24 °C: a) after inoculation, b) after cooling for 4.5 hours, c) after cooling for 22 hours.

A similar behavior was also observed during phase transition of other tin-drop contaminated MLM samples. Tin drops of Sn type D and E on the surface of MLM samples were generally found to break apart in less than a day, often in less than 12 hours, consistent with the data obtained for isolated drops shown in Fig. 3. Prior treatment in the central infected region of drops on MLM samples with HCl to remove tin oxides was applied only in a few cases. It was found not to be a necessary step for Sn of type D and E; infection with tin pest within a few hours could also be achieved by just spreading small α-Sn seed crystals on the untreated surface of the tin drop on the MLM sample. Obviously, mechanical deformation of tin deposits on MLM samples was not a viable option for increasing the cleaning speed due to high risk of optics damage.

The inherent deformation of the tin occurring during structural phase transition caused by the strong decrease in density is a clear advantage for cleaning purposes. It lowers the adhesion of the imbrittled droplet to the surface and causes buckling and lifting up of the created gray tin in the region of large α-Sn blisters, as seen for example in Fig. 4b. Eventually, this leads to fragmentation of the entire drop (Fig. 4c) enabling a facile subsequent contamination removal that is non-destructive for the coating, for example by jets of inert gas. Correspondingly, no damage of the coated sample surface could be observed in visual inspection with the microscope after removal of the created gray tin fragments by gas puffs of air. The gray tin fractured after transformation, since α-Sn regions are weaker and much less ductile compared to plastically deforming neighboring areas of β-Sn. Lattice distortions and internal stress led to the development of long cracks in the generated α-Sn material along slip lines parallel to the transformation propagation direction. Cracks developed also during growth when two regions of gray tin met (see Fig. 2). Observations under microscope after completion of the transition showed the grain size of the transformed tin and revealed a much finer polycrystalline microstructure with numerous cracks in comparison to the initial white tin.

It is known that the EUV reflectance of Mo/Si multilayer structures can degrade at elevated temperatures due to increased inter-diffusion occurring at the layer interfaces [37]. However, during dripping of molten tin with temperatures of ~240 °C onto the MLM sample its surface is affected by high temperatures only for a very short time. For comparison, detailed experiments and simulations for liquid tin drops with very similar size impacting on a steel plate at room temperature have demonstrated that the plate surface reaches its maximum temperature value in just a few milliseconds [38]. For the MLM samples used here, the duration for the surface temperature to return to near equilibrium after the heat transfer from the drop is estimated to be <0.5 s. For dripping of a single tin drop the time duration of elevated substrate temperatures is thus considered to be too short to cause any notable change in the reflectivity behavior of the coating. Prolonged cooling of the MLM to negative Celsius temperatures is also not expected to lead to any significant coating deterioration.

Nevertheless, potential optics degradation caused by the cleaning process of the tin drop was a concern requiring additional investigation. It was examined by exposing an uncapped MLM sample with previously measured EUV reflectance to the sequence of tin dripping, seed particle inoculation and β → α transformation at -24 °C, followed by removal of the converted gray tin. In this test an inoculated drop of tin type E was converted to gray tin fragments in just 7 hours after start of cooling. The EUV reflectance of the cleaned sample was then measured once more. The initial peak reflectance of the MLM coating was $R_{max}$ = 68.0 % at a wavelength of 13.61 nm. Its peak reflectance after cleaning, as determined by measurements at the PTB reflectometer [36], was $R_{max}$ = 67.0 % in the region covered by the tin drop, and in the range of 67.2 % - 67.7 % outside of it. No change in peak and center wavelength as well as half width was observed within measurement accuracy, thus confirming that the dripping of hot tin and the cleaning process by the phase transformation did not induce any significant damage in the MLM coating.

## 4. Conclusions

The tests carried out in this study have revealed the conditions for induction of tin pest on samples of smooth drops of β-Sn leading to their conversion into brittle pieces of α-Sn thus making them accessible for optics cleaning. It was found that the degree of purity of the material is an essential factor governing the speed of transformation of drops inoculated by seed particles. Prior deformation in addition to surface oxide removal and inoculation with α-Sn seeds was required for samples of tin purity grades lower than 99.99% in order to reach reasonably fast induction times of the phase transition during cooling at -24 °C. In contrast, for tin of 99.999% nominal purity, fast full structural phase transformation could be achieved in less than 24 hours of cooling by using only inoculation with α-Sn seed particles.

In a severe contamination scenario, smooth tin drops on MLM-coated EUV optics samples stored at -24 °C could be rendered cleanable in less than one day by disintegration initiated by induction of tin pest. It was shown that the EUV reflectivity of the MLM did not degrade substantially (at most by 1%) after tin drop transformation and removal. The basic concept of ex-situ optics cleaning by tin pest induction was thus proven in this work. Further studies are required to examine practicable possibilities of inducing the allotropic phase transition in the context of tin contamination removal, eventually also with in-situ techniques, in order to improve the light source availability in EUV lithography scanners. Specifically, the cleaning process based on tin pest induction can be optimized for multilayer-coated EUV collector modules and drive laser deflection mirrors.

## 5. Acknowledgements

ASML Netherlands B.V. is acknowledged for motivating this study. The experiments were carried out at BökoTech. I am grateful to the molecular and surface physics group of U. Heinzmann at Bielefeld University for general support. In particular, I thank Christian Meier (Bielefeld University) for helpful discussions and many valuable suggestions. Special thanks go to him and to Torsten Feigl (optiX fab) for kindly supplying MLM-coated EUV optics samples. Frank Scholze and his group at PTB Berlin are thanked for carrying out EUV reflectance measurements.

This research did not receive any specific grant from funding agencies in the public, commercial, or not-for-profit sectors.